\documentclass[letterpaper,twocolumn,10pt]{article}
\usepackage{usenix}

\usepackage[]{hyperref}

\usepackage{tikz}
\usepackage{amsmath}
\usepackage{xspace}
\usepackage{subfig}
\usepackage{booktabs}
\usepackage[frozencache,cachedir=./_minted]{minted}
\usepackage{comment}
\usepackage{multirow}
\usepackage{xurl}
\usepackage{tabularx}
\usepackage{pifont}

\def\Snospace~{\S{}}

\newcommand{\sys}{Phantora\xspace}

\newcommand{\paraspace}{\vspace{0.05in}}
\newcommand{\parab}[1]{\paraspace\noindent{\bf #1} }

\newcommand{\eg}{\textit{e.g.}}

\newcommand{\autorefsuffix}[2]{\hyperref[#1]{\autoref*{#1}#2}}

\begin{document}
\date{}

\title{\sys: Maximizing Code Reuse in Simulation-based Machine Learning System Performance Estimation} 

\author{
\textup{Jianxing Qin} \enskip \textup{Jingrong Chen} \enskip \textup{Xinhao Kong} \enskip \textup{Yongji Wu}$^\ddag$ \enskip \textup{Tianjun Yuan} \enskip \textup{Liang Luo}$^\dagger$
\\
\textup{Zhaodong Wang}$^\dagger$ \enskip \textup{Ying Zhang}$^\dagger$ \enskip \textup{Tingjun Chen} \enskip \textup{Alvin R. Lebeck} \enskip \textup{Danyang Zhuo}
\\
\\
\textit{Duke University} \enskip $^\dagger$\textit{Meta} \enskip $^\ddag$\textit{University of California, Berkeley}
} 

\maketitle
\begin{abstract}
Modern machine learning (ML) training workloads place substantial demands on both computational and communication resources. Consequently, accurate performance estimation has become increasingly critical for guiding system design decisions, such as the selection of parallelization strategies, cluster configurations, and hardware provisioning. Existing simulation-based performance estimation requires reimplementing the ML framework in a simulator, which demands significant manual effort and is hard to maintain as ML frameworks evolve rapidly.

This paper introduces \sys, a hybrid GPU cluster simulator designed for performance estimation of ML training workloads. \sys executes unmodified ML frameworks as is within a distributed, containerized environment. Each container emulates the behavior of a GPU server in a large-scale cluster, while \sys intercepts and simulates GPU- and communication-related operations to provide high-fidelity performance estimation. We call this approach hybrid simulation of ML systems, in contrast to traditional methods that simulate static workloads. The primary advantage of hybrid simulation is that it allows direct reuse of ML framework source code in simulation, avoiding the need for reimplementation. Our evaluation shows that \sys provides accuracy comparable to static workload simulation while supporting three state-of-the-art LLM training frameworks out-of-the-box. In addition, \sys operates on a single GPU, eliminating the need for the resource-intensive trace collection and workload extraction steps required by traditional trace-based simulators.
\sys is open-sourced at \url{https://github.com/QDelta/Phantora}. 

\end{abstract}
\section{Introduction}

Large machine learning (ML) models have become a driving force behind advancements in natural language processing~\cite{openai2024gpt4, touvron2023llama, vicuna2023}, computer vision~\cite{dosovitskiy2020visiontransformer}, computer graphics~\cite{openai2024dalle3} and recommendation systems~\cite{zhang2024wukongscalinglawlargescale,10.1145/3470496.3533727,MLSYS2024_78834433}. As models grow increasingly complex, high-performance model inference~\cite{kwon2023pagedattention} and training~\cite{zheng2022alpa} have become the main focus in the ML system community. Recent advancements in the field span from efficient GPU kernel optimizations~\cite{dao2022flashattention, dao2023flashattention2, kwon2023pagedattention}, to parallelization strategies~\cite{zheng2022alpa}, and scheduling algorithms~\cite{qiao2021pollux}. When deploying ML training jobs, it is often beneficial to estimate the system’s performance (\eg, training time per iteration, model FLOPS utilization), which can help operators decide how many hardware resources to allocate for a particular job, and plan for future hardware needs.

\begin{figure*}[t]
    \centering
    \includegraphics[width=.88\textwidth]{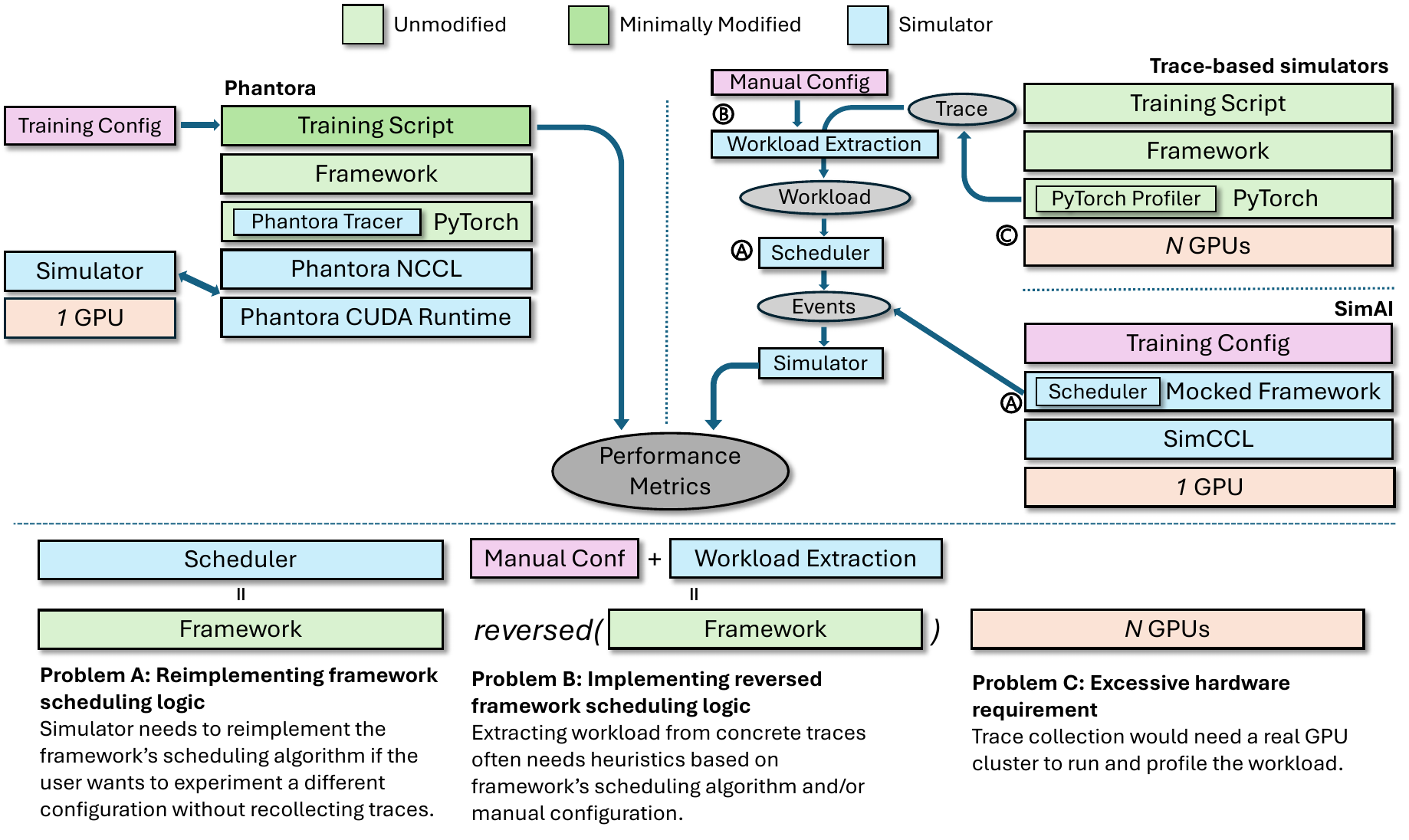}
    \caption{Comparison of simulators and problems of static workload simulation. Light green boxes show unmodified components; gree boxes show minimally modified components; blue boxes show simulator components and pink boxes show user input. Trace-based simulation requires workload extraction (reversing framework logic) and costly trace collection on large clusters. SimAI relies on mocked ML frameworks.}
    \label{fig:methods}
\end{figure*}

There has been a growing interest in performance estimation methods for ML systems. Analytical models (\eg, roofline \cite{roofline}) provide rapid estimates but lack accuracy. More recently, both industry and academia have shifted towards static workload simulation. \autoref{fig:methods} shows the two methods of static workload simulation and their problems. A common method is trace-based simulation~\cite{astrasim, astrasim2,daydream,dpro,flexflow,distsim,bang2024vtrain,gui2025multiverse}, where execution traces are collected from real runs of ML workloads on large clusters. These traces are then processed to extract workloads: 
a trace is lifted into higher levels of abstraction to make it suitable for configuration and event-driven simulation.  
Another method is SimAI~\cite{wang2025simai}, which implements a mocked version of ML frameworks in order to generate events that can be used in a simulator. However, both methods have to reimplement the scheduling logic of the simulated ML framework (\eg, DeepSpeed~\cite{rasley2020deepspeed}, Megatron~\cite{shoeybi2019megatron}, TorchTitan~\cite{liang2025torchtitan}) and trace-based simulation may even need to implement a reversed scheduling logic in workload extraction.
Reimplementation makes them difficult to maintain due to fast-evolving ML training frameworks.

In this paper, we explore the following question: \textit{Can ML framework source code be directly reused in simulation-based performance estimation?}
Modern ML frameworks already provide a rich set of configurable parallelization strategies (\eg, pipeline parallelism, expert parallelism, data parallelism) and rematerialization strategies (\eg, ZeRO~\cite{rajbhandari2020zero}, FSDP~\cite{zhao2023pytorchfsdpexperiencesscaling}, activation recomputation~\cite{korthikanti2023ac}). By reusing these implementations, we can evaluate different parallelization and rematerialization strategies without relying on the simulator’s reimplementations. In addition, the performance benchmarking code embedded in training scripts can be reused directly, allowing users to interact with the simulator in the same way they would when tuning ML system performance on a real GPU cluster. Most importantly, avoiding reimplementation of the framework greatly reduces the maintenance burden, making the simulator more sustainable as ML frameworks evolve rapidly.

We present a fundamentally different performance estimation approach for ML systems, called \textit{hybrid simulation}. Our key idea is that we can \textit{integrate real system execution directly with event-driven simulation, creating an illusion for ML frameworks that they are running on a real GPU cluster.} 
Based on this approach, we build \sys, a GPU cluster simulator for ML system performance estimation. The core design is to use a containerized environment, equipped with a single GPU to directly run an ML system for simulation. Each running container simulates the execution of a multi-GPU server, where each rank operates one simulated GPU and is provided with one virtual clock. \sys intercepts all the communications and GPU kernel invocations from the ML framework. CUDA kernel execution times are profiled on the single GPU, while communication execution times are calculated using an event-driven network simulator. When a rank launches a CUDA kernel or initiates communication, \sys adjusts the rank's virtual clock accordingly to maintain accurate simulated time.

We still face three key research challenges. First, it is challenging to maintain the core abstractions (e.g., CUDA, NCCL) so that unmodified ML frameworks can directly run. Second, it is challenging to integrate a real, running system with an event-driven network simulator. To elaborate on the second challenge, an event-driven network simulator progresses in discrete time steps, and the timing for the next event is calculated by prior events in the system. However, in hybrid simulation, our distributed ML system (the set of containers) may inject a network flow at any time, causing the network simulator's calculation to be incorrect.
Finally, certain distributed ML workloads may quickly exhaust the memory capacity of the simulation environment.

\sys maintains the core abstractions used by PyTorch and uses runtime patching to dynamically rewrite their dependencies (e.g., timer), so that the ML frameworks (e.g., Megatron \cite{shoeybi2019megatron}, DeepSpeed \cite{rasley2020deepspeed}, TorchTitan \cite{liang2025torchtitan}) can run as is on top of \sys.
\sys addresses the integration challenge by allowing an event-driven network simulator and a distributed ML system to run in a loosely synchronized manner. 
When an event that should occur earlier is injected by the container to the network simulator, \sys rollbacks the simulator state and accommodates those past events triggered by the containers. 
To enhance scalability, \sys allows containers to share memory, significantly reducing the simulator's memory footprint.
We evaluate \sys using three state-of-the-art Large Language Model (LLM) training systems: Megatron~\cite{shoeybi2019megatron}, DeepSpeed~\cite{rasley2020deepspeed}, and TorchTitan~\cite{liang2025torchtitan}. 
All three systems run out-of-the-box without any modification to the source code, and \emph{\sys can support their feature sets without the need for corresponding reimplementation.}
Their terminal outputs remain identical (except training losses) to those produced when running on a real GPU cluster.
Our small-scale NVIDIA H200 testbed evaluation demonstrates that \sys achieves simulation accuracy comparable to state-of-the-art workload simulation methods. \sys's simulation result on TorchTitan matches its reported performance on large-scale NVIDIA H100 and A100 clusters~\cite{liang2025torchtitan,torchtitana100}.

This paper makes the following contributions:
\begin{itemize}
\item We are the first to propose \textit{hybrid GPU cluster simulation for ML systems}, which eliminates the need for reimplementing ML frameworks in the simulator.

\item We design and implement \sys, which efficiently integrates real system execution with an event-driven simulation to enable hybrid simulation.

\item We evaluate \sys on three state-of-the-art LLM training systems, demonstrating \sys's generality and accuracy.
\end{itemize}
\section{Background}

Performance estimation is valuable for both ML system developers and infrastructure providers. For system developers, it enables rapid evaluation of the performance of the systems they are building. As model sizes grow, many system parameters require careful tuning, for instance, selecting an appropriate parallelization strategy~\cite{shoeybi2019megatron, zheng2022alpa}. Being able to estimate the performance of different strategies makes it easier to identify the most efficient option. For infrastructure providers, performance estimation allows planning for future hardware deployments.

Due to the accuracy limitations of performance modeling (e.g., roofline \cite{roofline}), developers have turned to static workload simulation for performance estimation. 
Today, two workload simulation methods exist.
The first method is trace-based simulation~\cite{astrasim, astrasim2,daydream,dpro,flexflow,distsim,bang2024vtrain,gui2025multiverse}. A trace is collected in a real execution, and a workload can be extracted from a collected trace. Another method is SimAI~\cite{wang2025simai}, where workloads are generated via implementing a mocked version of the ML frameworks.
Workload simulation leads to accurate performance estimation and is increasingly adopted by industry~\cite{sridharan2023chakra}. 

\begin{figure}[t]
    \centering
    \includegraphics[width=.47\textwidth]{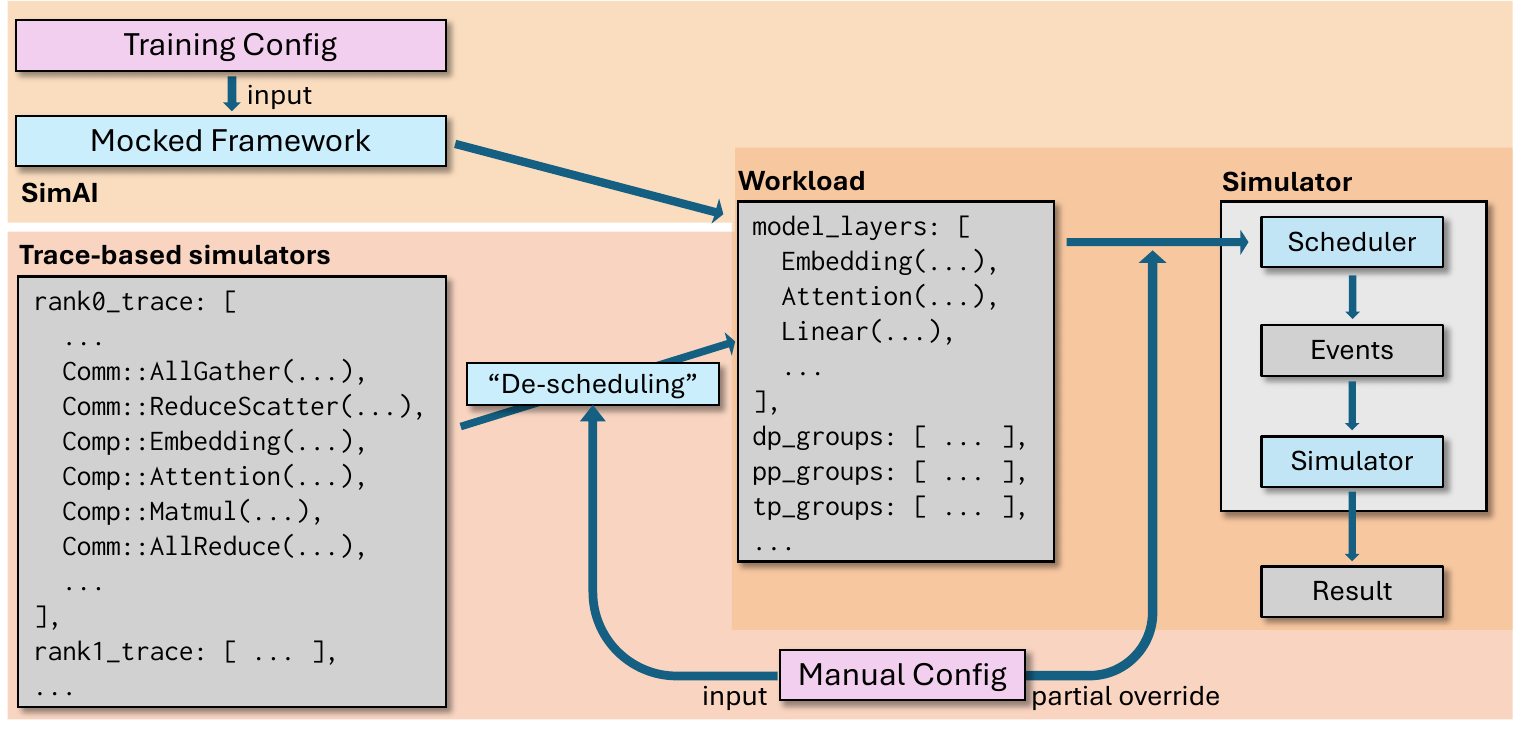}
    \caption{Scheduling logic needs to be reimplemented in current simulators for converting high level workload to detailed computation and communication events.}
    \vspace{-4mm}
    \label{fig:background}
\end{figure}

\parab{Why do static workload simulators need to reimplement ML frameworks?}
For trace-based simulators, there are two fundamental reasons why reimplementation of the ML frameworks is necessary.
First, the key goal of simulation is to explore alternative configurations. Yet, a collected trace reflects only the specific configuration used during execution. To enable configuration exploration, trace-based simulation requires a workload extraction step that lifts the trace into abstract workload, revealing higher-level configurations from actual traces (\autoref{fig:background}). This extraction step typically relies on human understanding of ML framework execution, effectively constructing a reversed version of the framework’s logic. Heuristics can be brittle and may fail to generalize across frameworks; therefore, trace-based simulation often relies on extra manual configurations to help with workload extraction, which increases manual effort.
More importantly, after the user changes the configuration in abstract workload, the simulator needs to turn user's configurations into a detailed execution plan (events) for actual simulation. This scheduling also needs to correctly reflect the framework's scheduling being simulated, which is essentially a reimplementation of the framework's logic.

Beyond trace-based simulation, SimAI \cite{wang2025simai} adopts an alternative approach. Rather than extracting workloads from collected traces, it uses mocked frameworks to directly produce low-level events. However, while this removes the burden of workload extraction, reimplementing scheduling remains necessary in the mocked frameworks. It tightly couples SimAI to specific framework and versions: whenever the underlying ML framework evolves or new framework appears, SimAI’s logic must be updated accordingly to ensure correctness. 

Due to the need of reimplemeting frameworks, both trace-based simulators and SimAI struggle to provide complete support of features in current frameworks and new frameworks. For example, none of the existing simulators support TorchTitan~\cite{liang2025torchtitan}, and none of them can analyze throughput and memory usage of selective activation checkpointing~\cite{korthikanti2023ac,liang2025torchtitan} at the same time. Simulators may even fail to generate the exact same model as the framework. For example, when given the same configuration matching Llama2 7B \cite{touvron2023llama}, the size of model generated by SimAI differs by 7.4\% from native \texttt{GPTModel} in Megatron \cite{shoeybi2019megatron} model library.

\parab{Our approach: Hybrid simulation.} Our objective is straightforward: to build a general simulator in which ML frameworks’ source code can be directly reused. Modern ML frameworks already provide well-maintained implementations and highly configurable parallelization strategies. Leveraging them directly makes the simulator both easier to maintain and more broadly applicable across different frameworks. To achieve this, our approach is to construct a GPU cluster execution environment that closely mirrors a real one, allowing ML frameworks to run on top of the simulated cluster with no modification. Next, we discuss the research challenges to realize this approach and the overview of our design.

\section{Overview}

We face three key research challenges in realizing hybrid simulation of ML frameworks: (1) supporting unmodified ML frameworks, (2) ensuring correct and efficient time synchronization, and (3) achieving scalability.
First, modern ML frameworks have complex software and hardware dependencies. They assume the presence of GPUs, NVLinks, and RDMA networks, and they rely on a wide range of libraries. Constructing an execution environment in which an ML framework can run unmodified is therefore nontrivial.
Second, integrating event-driven network simulation with ML framework execution presents a fundamental mismatch. Network simulators advance time in discrete time steps, while ML frameworks execute with continuous, real-time progression. Reconciling these two timing models is challenging.
Finally, ML frameworks are resource-intensive, consuming substantial compute, memory, and communication bandwidth. While we would like to execute them as-is, we must avoid the prohibitive resource costs of training a full-scale model on a real cluster.

\begin{figure}[t]
    \centering
    \includegraphics[width=.48\textwidth]{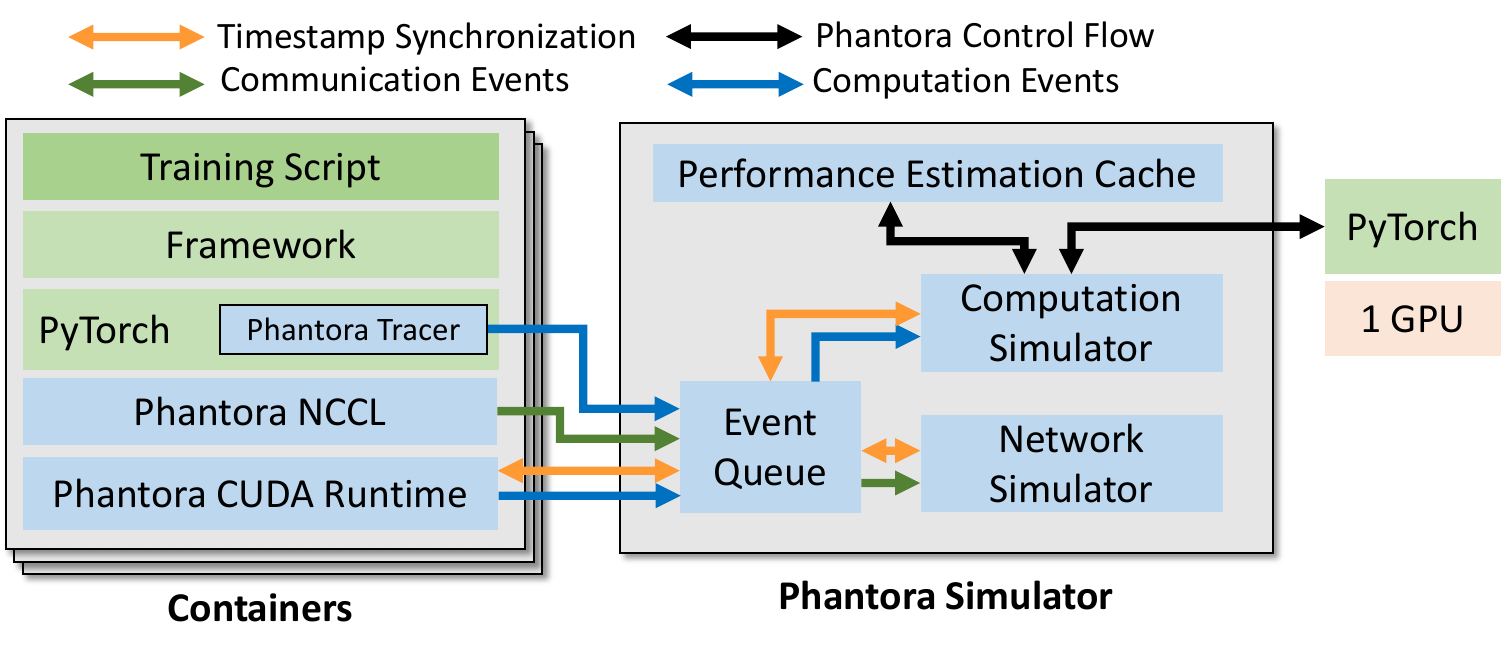}
    \caption{\sys architecture. Components in light green are unmodified, components in green are minimally modified, and components in blue are constructed by \sys.}
    \vspace{-2mm}
    \label{fig:overview}
\end{figure}

To support unmodified ML frameworks, \sys runs an ML framework in a realistic containerized environment and interacts with this real system for simulation. \autoref{fig:overview} shows \sys's architecture. Each container in the environment acts as a GPU server.
Each container runs Python interpreters that execute ML framework code.
There are two operations that require interaction between the containers and the \sys simulator: (1) GPU computation and (2) communication. When a computation kernel is invoked, \sys uses a single GPU to profile the kernel's execution time. For communication, \sys utilizes an event-driven flow-level network simulator to estimate completion time.
\sys only needs a single GPU because it profiles the performance once for each (computation kernel, tensor shapes) combination. This is sufficient because computation kernel performance is usually independent of the tensor values.\footnote{There are some exceptions. One such example is sorting where conditional branches could be chosen based on comparisons. We discuss this point in \autoref{sec:discussion}.}
In this way, an application cannot distinguish whether it is running on \sys or a physical GPU cluster as long as its control flow does not depend on tensor values (which would be junk values in \sys). The time of each rank in every container will be maintained by \sys using standard discrete event simulation, and the application can read this time to calculate its performance. A naive implementation of the above approach would still require modifying the ML framework source code. For example, one might need to change the performance timer implementation used in framework logging. To support ML frameworks out of the box, we instead leverage Python’s runtime patching to dynamically redirect underlying dependencies of an ML framework.

\begin{figure*}[t]
    \centering
    \includegraphics[width=.90\textwidth]{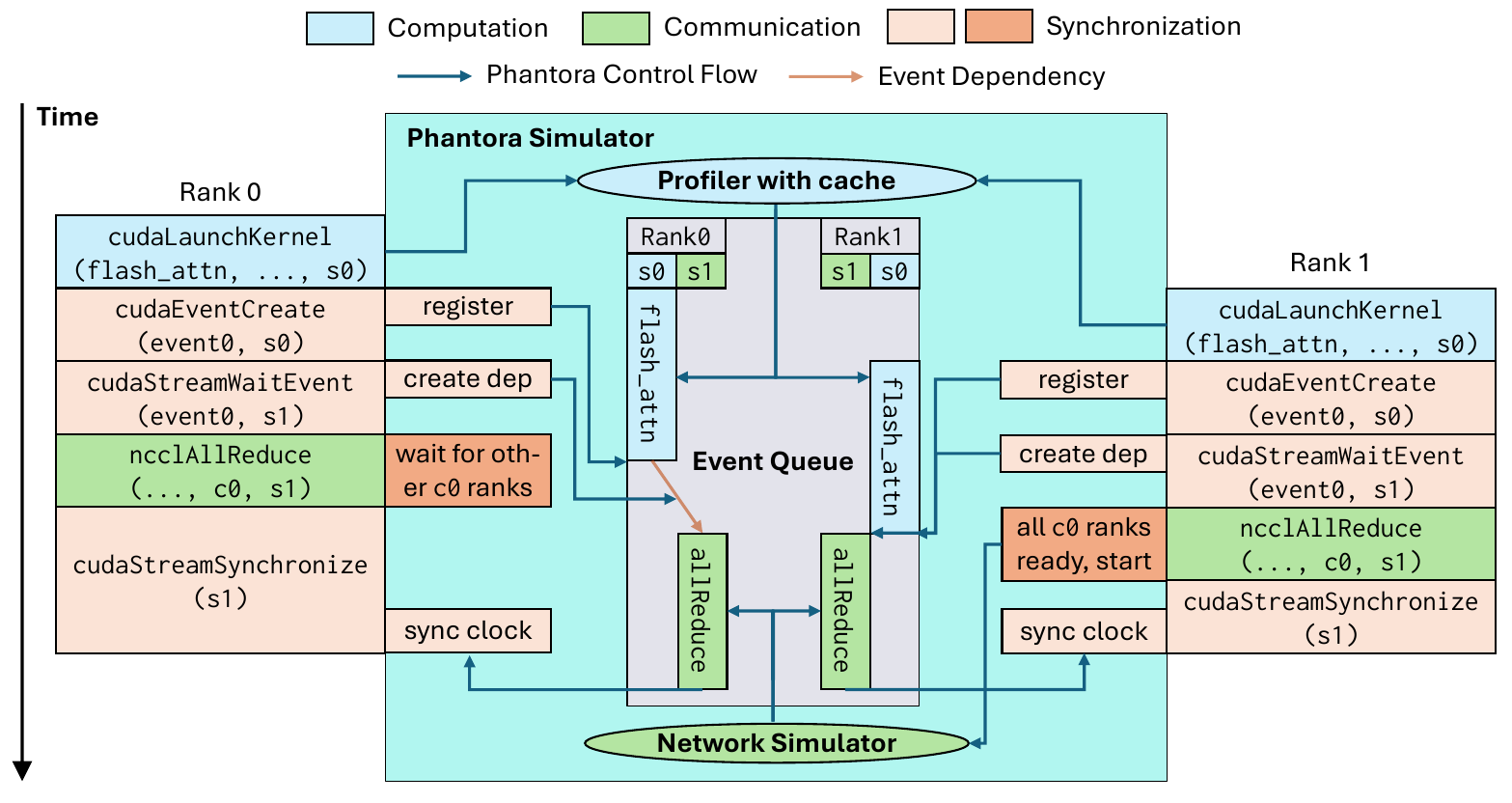}
    \caption{An example workflow of \sys with two ranks. ML system places computation and communications on different CUDA streams for flexibility, and use CUDA events to manage the synchronization between them. \sys needs to correctly handle these synchronizations to achieve accurate simulation.}
    \vspace{-3mm}
    \label{fig:workflow}
\end{figure*}

To appreciate the time synchronization challenge, let's consider the following \textit{past events} scenario.
Traditional event-driven simulators require static workloads. For example, in a typical event-driven simulation workflow, all the events in the workload are pre-loaded into the event queue of the simulator. The simulator processes these events in chronological order, updating the simulation state and queuing new events as necessary. However, with real systems, events are generated dynamically and injected into the simulator, which can lead to past event scenarios: the simulator receives an event generated by the real system after the simulator has already advanced to the next event, which has a timestamp later than the real system's event.
One solution is to keep the event-driven network simulator tightly synchronized with the containers. In hardware simulators, for example, it is common to use a tiny time quantum and synchronize all components at every quantum~\cite{reinhardt1993wwt,mukherjee2000wwt}. However, it introduces significant overhead and defeats the purpose of event-driven simulation in modern network simulators.
\sys uses loose time synchronization between the event-driven simulator and the real system execution to address this issue and enables fast simulation. To ensure simulation accuracy, we implement an event-driven network simulator capable of time traveling to rollback the states of all flows when events occur in the past.

We apply two techniques to scale \sys. In LLM training, each GPU server may load training data and model weights, which consumes significant host memory. Our evaluation (\autoref{sec:eval-scalability}) shows that a physical server with 256GB host memory can only simulate 9 GPUs for the training of Llama2 7B model without our technique. We introduce a novel memory sharing mechanism for different containers running on the same host, which significantly reduces the memory footprint. 
Second, \sys measures only the actual CPU time consumed by each process, rather than the wall clock time. This choice preserves simulation accuracy in the presence of CPU core contention, where wall-clock measurements would otherwise overestimate execution time due to frequent context switches between the containers.

\section{Details of \sys}

\subsection{Supporting Unmodified ML Frameworks}
\label{sec:workflow}

Each rank executes unmodified ML framework code, using PyTorch and NCCL libraries, and interacts with the CUDA Runtime. We implement the \sys Tracer in the PyTorch. This tracer collects all invoked PyTorch operators and the corresponding performance-related parameters, and pushes all information as computation events to the event queues in the \sys simulator. It is worth noting that this tracer does not affect the execution of PyTorch, and the operators are still dispatched to the corresponding CUDA backend to initiate computation in the CUDA Runtime. We replace the native CUDA Runtime with \sys CUDA Runtime, which does not actually execute any CUDA calls. Instead, it only maintains necessary metadata to emulate actual CUDA Runtime behaviors. For example, \texttt{cudaMalloc}/\texttt{cudaFree} in \sys does not actually allocate/deallocate GPU memory, but only tracks GPU memory usage and returns \texttt{cudaErrorMemoryAllocation} when an allocation will make usage exceed the configured memory capacity. In addition, it pushes CUDA calls as events to the event queue in \sys simulator. Similarly, we replace the native NCCL library with \sys NCCL library. \sys NCCL library does not initiate communication, but forward all communication operations (\eg, allreduce) to the simulator by pushing communication events to the event queues. Furthermore, \sys maintains a dependency graph of events to emulate CUDA's asynchronous semantics, i.e. CPU launches computation and communication kernels and specify dependencies through streams/events.

\autoref{fig:workflow} shows an example workflow of \sys and two ranks. These two ranks are running a distributed workload of attention and all-reduce on the attention result. The launch of FlashAttention~\cite{dao2022flashattention,dao2023flashattention2} is captured by \sys CUDA Runtime and pushed to \sys simulator. PyTorch operators traced by \sys Tracer are also pushed to \sys simulator as computation events. \sys simulator will then invoke native computation libraries to profile these computation kernels. The profiler uses a performance estimation cache to store the performance results of operators that have been already faithfully executed. When invoking the same operators in the future, \sys will directly use results stored in the cache. In this example, the FlashAttention call of Rank 1 will not be profiled again, instead, \sys simulator will use the cached profiling result of Rank 0's FlashAttention.

To accurately capture the dependencies and synchronizations of computation and communications and emulate the behavior of actual CUDA Runtime, \sys needs to carefully handle related CUDA calls. \sys event queue is designed to natively support dependencies and is used to emulate CUDA streams and events--two core constructs in CUDA asynchronous programming. Operations on the same stream will have implicit dependency in chronological order, and operations on different streams have no dependency unless explicitly specified via CUDA events. In \autoref{fig:workflow}'s example, computation and communications are launched on different streams for flexibility, and the dependency between them is enforced via an additional CUDA event. \sys correctly emulated that via dependencies in the event queue.

\sys NCCL captures communication operations and pushes them to the \sys simulator. For example in \autoref{fig:workflow}, \texttt{ncclAllReduce} is first called by Rank 0. This is API is non-blocking so \sys NCCL will return immediately after pushing the call to the simulator, but the simulator will not start network flows until all ranks in the same communicator are prepared (in this example, wait for Rank 1 to call \texttt{ncclAllReduce} with \texttt{c0}), which is in compliance with NCCL semantics.
To accurately model the execution time of collective communication operations, \sys adopts a standard flow-level network simulator adapted from NetHint~\cite{chen2022nethint}, referred to as \texttt{netsim}. The \texttt{netsim} simulator takes a cluster topology configuration as input, where users can specify various properties of the cluster, including switch port bandwidth, cluster interconnection, and multipath routing and load balancing strategies. The throughput of flows at each time is computed based on the max-min fairness. Within \texttt{netsim}, we implement the communication patterns of different collective operations. For instance, we model \texttt{allreduce} using a ring-based approach, as configured in NCCL in our evaluation. When receiving communication operators, \sys submits the corresponding data transfer flows to \texttt{netsim} with the appropriate timestamps. \texttt{netsim} then simulates network behaviors and computes the completion time of each flow based on network congestion and available bandwidth~\cite{chen2022nethint, wu2024mccs}.

Time synchronization is enforced through CUDA synchronization calls, which should block the host until certain GPU completion point (\eg, \texttt{cudaStreamSynchronize}).
When these APIs are called, the \sys CUDA Runtime pushes a synchronization event to the event queue and starts waiting for a response. After processing the preceding events in the event queue and completing this synchronization event, the simulator returns a response, including a completion time (a logical timestamp), to the rank's \sys CUDA Runtime. The rank's virtual clock is then updated based on this completion time. In the example shown in \autoref{fig:workflow}, The virtual clock of both Rank 0 and Rank 1 will be updated to the completion time of all-reduce after \texttt{cudaStreamSynchronize}.

\parab{Runtime patching for ML frameworks.}
For certain frameworks, there might be unconfigurable behaviors that do not satisfy \sys's assumption. To ensure no modification to the framework code and provide a close-to-natural user experience, we leverage the dynamic features of Python to runtime patch certain functions. For example, the built-in performance logging of TorchTitan utilizes \texttt{time.perf\_counter} that needs to be replaced by \sys timers. With runtime patching, \sys can directly work with ML frameworks installed via ``\texttt{pip install}''.

\parab{Intercepting CUDA kernel invocations and communication.}
One key design decision we have to make is how to intercept CUDA kernel invocations and communications. A strawman solution could intercept ML system execution only at the CUDA kernel level, which is widely used in profilers such as Nsight Systems~\cite{nsight_systems}.
But at this level only untyped pointers to arguments are provided, so the inspection of arguments would require extra configurations for each kernel.

Hence, we resort to a hybrid approach where most computation operations are intercepted at ML systems API level (\eg, PyTorch operators), while communication operators and some specific computation operations like FlashAttention~\cite{dao2022flashattention,dao2023flashattention2} are intercepted at runtime libraries level (\eg, NCCL, CUDA Runtime). This hybrid approach allows \sys to maintain generalizability without introducing the development effort of inspecting every CUDA kernel involved. For computation operations, \sys is aware of the type of operations and the shapes of the input tensors. \sys implements a cache manager to reuse earlier profiling results of an operation with the same input shapes, eliminating redundant execution.

\subsection{Loose Time Synchronization of Real Execution and Event-Driven Simulation}
\label{sec:integration}

\begin{figure}[t]
    \centering
    \includegraphics[width=.48\textwidth]{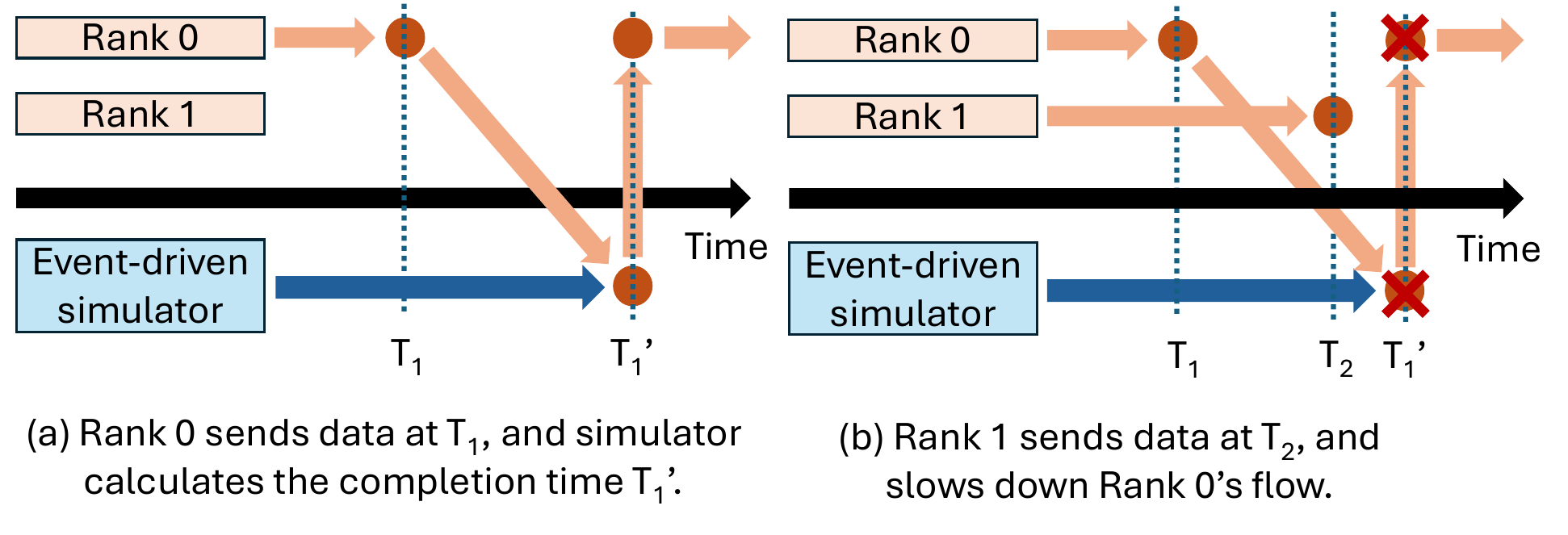}
    \caption{Challenges of synchronizing time between real execution and event-driven simulation.}
    \vspace{-3mm}
    \label{fig:synchronization}
\end{figure}

\sys needs to correctly synchronize the real execution and the event-driven simulator for accurate end-to-end simulation, as events generated by the real execution may impact the event-driven simulator. Without proper synchronization, such impacts may be neglected or wrongly considered, leading to inaccurate simulation. 

\autoref{fig:synchronization} demonstrates a typical workload that requires correct synchronization. 
Rank 0 and 1 independently launch communications that may share network resources. At time $T_1$, Rank 0 begins to wait for its communication and asks the simulator for the completion time. The simulator, however, cannot directly proceed to the completion point based on current information and respond $T_1'$, as a future communication from Rank 1 starts at $T_2$ may affect this completion time if $T_2 < T_1'$. Note that this is not an issue if the simulated workload is static (\eg, the use cases of today's static workload simulators). For static workload, the simulator already knows the existence of event at $T_2$, so it can simply move the simulated time forward to $T_2$ instead of $T_1'$, and update the state of the system. However, in our case because Rank 1 is a real execution, the simulator does not know whether or when it will launch communication.

One approach to resolve this issue is to determine a small time quantum, and move time forward in both real execution and simulation. This is a common technique in hardware simulation. WWT~\cite{reinhardt1993wwt} uses such an approach for simulating cache-coherent shared memory multiprocessors, and it determines the time quantum based on cross-core communication latency. However, using a fine-grained time quantum can significantly slow down the simulation speed, which is exactly why it is not used by most of today's network simulators.

Our key observation is that the characteristics of ML systems provide an opportunity to change the running time of operations during hybrid simulation. Specifically, the time taken by a given operation (\eg, a specific communication) does not affect the actual control flow of the real system. For example, changing the time consumed by a matrix multiplication in a rank's runtime does not affect which is the next operator invoked by the rank. Therefore, we can \emph{rollback} the simulator state and correct the real system state efficiently during simulation.
We apply this insight by optimistically synchronizing clocks between the simulator and each rank's runtime in \sys. When past events occur, the simulator rollbacks to a prior time, processes the past events, corrects time for a set of events in the past, updates the clock with each rank's runtime, and continues the simulation process.

\begin{figure}[t]
    \centering
    \includegraphics[width=.48\textwidth]{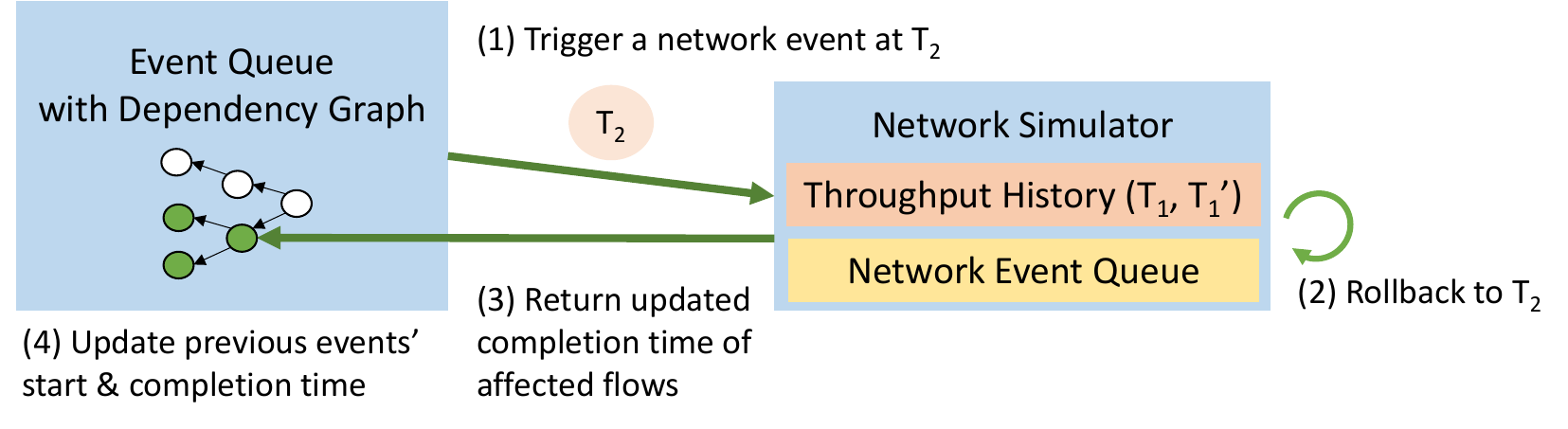}
    \caption{Handing past events in the event-driven simulator using time rollback. The network state at $T_2$ is a superposition of the states at $T_1$ and $T_1'$.}
    \vspace{-2mm}
    \label{fig:rollback}
\end{figure}

\parab{Time rollback.}
A traditional event-driven simulator keeps a priority queue of ``events'', where the priority is the start time of the event. This allows the simulator to process the event in the chronological order. \sys augments this simulator by adding the ability to time travel to any particular time in the past. To realize this feature, the network simulator keeps the throughput history of all flows. Consider the same example in \autoref{fig:synchronization}. A new event arrives at a time $T_2$ earlier than the current time $T_1'$. For simplicity, let's assume there are no other events between $T_1$ and $T_1'$ and let $S(T)$ denote the state of all the network flows at time $T$. As \autoref{fig:rollback} shows, the network simulator can compute state $S(T_2)$ based on the stored throughput history between $S(T_1)$ and $S(T_1')$.  This is because between neighboring events, network flows are assumed to have stable throughput, which is a common assumption made by existing event-driven network simulators. Such a time rollback can affect previously computed completion time of some network flows. The network simulator then sends these updated completion times to \sys's event queue to update other events' start/completion time by traversing the dependency graph. For example, if another event previously started at $T_1'$ for its dependency on this communication, then its start and completion time may be adjusted accordingly.

With this rollback mechanism, \sys will still ensure forward progress if the system being simulated ensures forward progress.
In \sys, the simulated ML system only has finite past events, so there will only be finite rollbacks and the simulator will make forward progress after rollbacks are complete.

\parab{Garbage collection of historical states.} The ability to time travel to the past comes with the cost of storing the simulation states at all the event timestamps. These states include the dependency graph stored in \sys's event queue and the historical flow states in the network simulator. As the simulation progresses, storing these historical states can occupy a lot of host memory. \sys implemented garbage collection to address this issue. The key insight is that after all the ranks' time has passed $T$, it is impossible to inject an event before $T$ into the network simulator. Thus, all the simulator states before $T$ (including both the dependency graph and the flow states) can be safely discarded.

\parab{Network simulator with time rollback.} Our flow-level simulator enables time rollback with low additional runtime overhead. The simulator assumes per-flow fairness across the network and solves the max-min fair flow allocation problem using an iterative water-filling algorithm. At each iteration, the simulator identifies the bottleneck link and computes the necessary delta adjustments for flow rates.

To support time rollback, we provide two APIs: one for updating the start time of an existing flow, and another for advancing the simulation by one step or up to a specified time. The simulator records the throughput history for each flow, which is represented by a few floating point numbers. Since throughput changes are regular events to a network simulator, without garbage collection, the memory overhead is proportional to the number of discrete events the simulator processes.
Although a rollback can potentially update multiple flows, these computations are based solely on the throughput history and can be computed in an incremental manner, making them computationally efficient compared to solving the max-min fair flow allocation problem.

\subsection{Improving \sys's Scalability}
\label{sec:scalability}

\sys has the following two techniques to achieve scalable simulation.

\parab{Scalability Technique \#1: Model parameter sharing on CPU.} ML systems may initialize models in CPU memory, either randomly or using pre-trained weights stored on disk. Once initialized, the models are transferred to the GPU for subsequent tasks. In real GPU clusters, this initialization phase is generally not a bottleneck, as GPU memory size is usually smaller than the CPU memory size. However, when \sys simulates a large cluster using limited hardware resource, this peak memory usage could become a scalability constraint.
To address this limitation, \sys implements parameter sharing, which allows model parameters on the same simulation server to be transparently mapped to the the same region of shared memory.
This ensures that at most one copy of the model is initialized per server. \sys assumes that the control logic of ML systems does not depend on tensor values, allowing safe sharing of model parameters without impacting execution.

\parab{Scalability Technique \#2: Use CPU time instead of system time.} Similar to memory, CPU can also become a bottleneck for \sys. Existing ML training systems are usually multi-process. If the number of CPU cores used for simulation is smaller than the number of processes launched, CPU oversubscription can slow down the execution of the ML systems and cause inaccuracies in simulation results. To address this problem, \sys only counts the actual CPU time each process spent instead of the system time passed (wall clock). Thus, although the simulation process is still slowed down, the accuracy of the results will not be affected. \sys can also be configured to ignore the CPU time completely, leaving only the GPU operation time and CUDA synchronization waiting time to be included in the results.

\section{Evaluation}
\label{sec:eval}
Our \sys prototype consists of 9K lines of Rust, 1.8K lines of C, and 500 lines of C++: 1.8K lines of C and 1K lines of Rust for \sys NCCL and CUDA Runtime, 3.6K lines of Rust for the flow-level network simulator, 3.4K lines of Rust for the event queue, 1K lines of Rust for the computation simulator, and 500 lines of C++ for \sys Tracer.

We evaluate \sys on three aspects. First, we test \sys's generality in supporting different ML frameworks, and their runtime behaviors. 
Second, we test a set of standard metrics for simulators, such as accuracy, simulation speed and scalability. Finally, we do a case study on selective activation recomputation \cite{korthikanti2023ac} to show \sys's capability on estimating ML system performance and GPU memory usage for features that existing static workload simulators have not fully reimplemented.

\subsection{Generalizability}
\label{sec:generalizability}

\parab{Effort for supporting ML frameworks.} \sys currently support three LLM training frameworks: Megatron~\cite{shoeybi2019megatron}, DeepSpeed~\cite{rasley2020deepspeed} and TorchTitan~\cite{liang2025torchtitan}.
Users can simply \texttt{pip install} these frameworks from official PyPI without building from source or using a different package index. All the runtime patches are applied when users \texttt{import} our helper library. Specifically, the size of runtime patches of these three frameworks are: \emph{Megatron}: no patch needed. \emph{DeepSpeed}: 4 lines of code where a NCCL setup validation is disabled. \emph{TorchTitan}: 1 line of code where \texttt{time.perf\_counter} is replaced with \sys timer.

We believe supporting other PyTorch based frameworks in \sys should also require minimal effort. Further, \sys does not depend on model architecture. Our evaluation results focus on LLMs. See \autoref{sec:nonllm} for our evaluations using non-LLM workloads.

\parab{Modifications to the training script.} For every training script, \sys Tracer needs to be explicitly enabled at the beginning and disabled at the end. This, together with importing of our helper library, adds about 6 lines of extra code per training script.

In addition, when running Megatron \cite{shoeybi2019megatron} with \sys, gradient clipping must be disabled. This feature performs fallible CPU operation (specifically, square root) of data copied from GPU, which could lead to math errors as GPU memory value is effectively random in \sys.

\parab{User experience of \sys.}
With \sys, users are able to tune their ML system performance as if they are actually experimenting with a real GPU cluster. \sys natively supports customized printing and logging mechanisms embedded in the ML systems.

\begin{figure}[t]
\begin{minted}[fontsize=\footnotesize]{python}
time_delta = timer() - time_last_log
# tokens per second, abbr. as wps by convention
wps = ntokens_since_last_log / (
    time_delta * parallel_dims.model_parallel_size
)
# model FLOPS utilization
mfu = 100 * num_flop_per_token * wps / gpu_peak_flops
time_end_to_end = \
    time_delta / job_config.metrics.log_freq
time_data_loading = np.mean(data_loading_times)
metrics = {
    "wps": wps,
    "mfu(%)": mfu,
    "time_metrics/end_to_end(s)": time_end_to_end,
    "time_metrics/data_loading(s)": time_data_loading,
    ...
}
metric_logger.log(metrics, step=train_state.step)
logger.info(
    f"step: {train_state.step:2}  "
    f"loss: {global_avg_loss:7.4f}  "
    f"memory: {gpu_mem_stats.max_reserved_gib:5.2f}GiB"
    f"({gpu_mem_stats.max_reserved_pct:.2f}%)  "
    f"wps: {round(wps):,}  "
    f"mfu: {mfu:.2f}%{color.reset}"
    ...
)
\end{minted}
\subfloat{
    \centering
    \includegraphics[width=.48\textwidth]{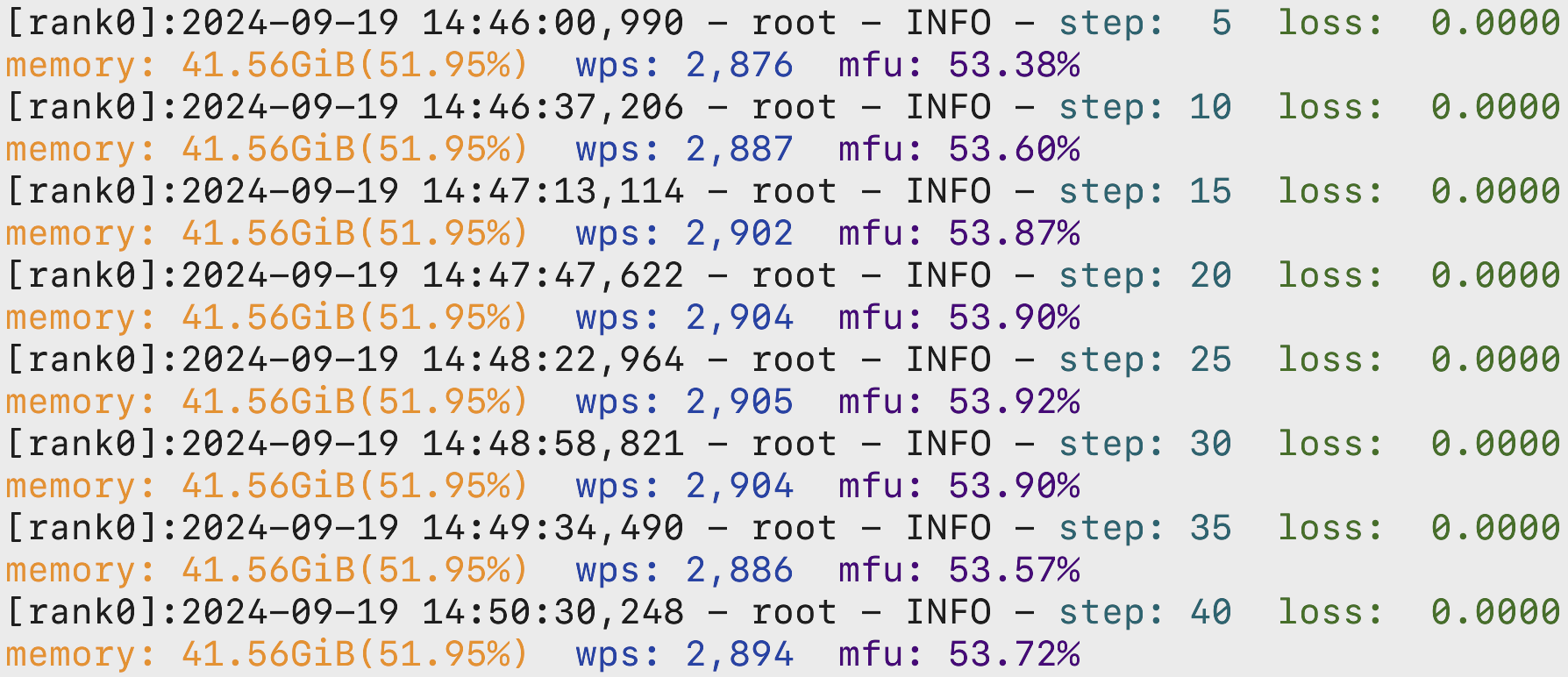}
}
\caption{Performance estimation code in TorchTitan~\cite{torchtitantrain} and the console output of running TorchTitan on top of \sys. The console output is exactly the same as if TorchTitan runs on a real GPU cluster except losses.}
\vspace{-1mm}
\label{fig:measurement-code}
\end{figure}

The top part of \autoref{fig:measurement-code} shows the performance measurement and logging code in TorchTitan~\cite{liang2025torchtitan}, which represents how TorchTitan developers want to evaluate performance. In other existing simulators, printing these metrics would require a reimplementation of this code in post-simulation analysis. In contrast, \sys allows this code to run as is, and users can see results in exactly the same format as if they actually run the ML system on a real GPU cluster. The bottom part of \autoref{fig:measurement-code} shows the console output of running TorchTitan on top of \sys. To the best of our knowledge, \sys is the only method that has this type of generalizability. 

After developers update either the ML system code (\eg, changing parallelization strategies for LLM training), the model or the performance measurement and logging code, \sys can immediately re-run and produce updated console output. For other existing simulators, the developer may have to craft extra configurations, collect additional traces or change post-simulation analysis---all of which significantly slow down the development cycle.

\begin{figure}[t]
\centering
\includegraphics[width=.48\textwidth]{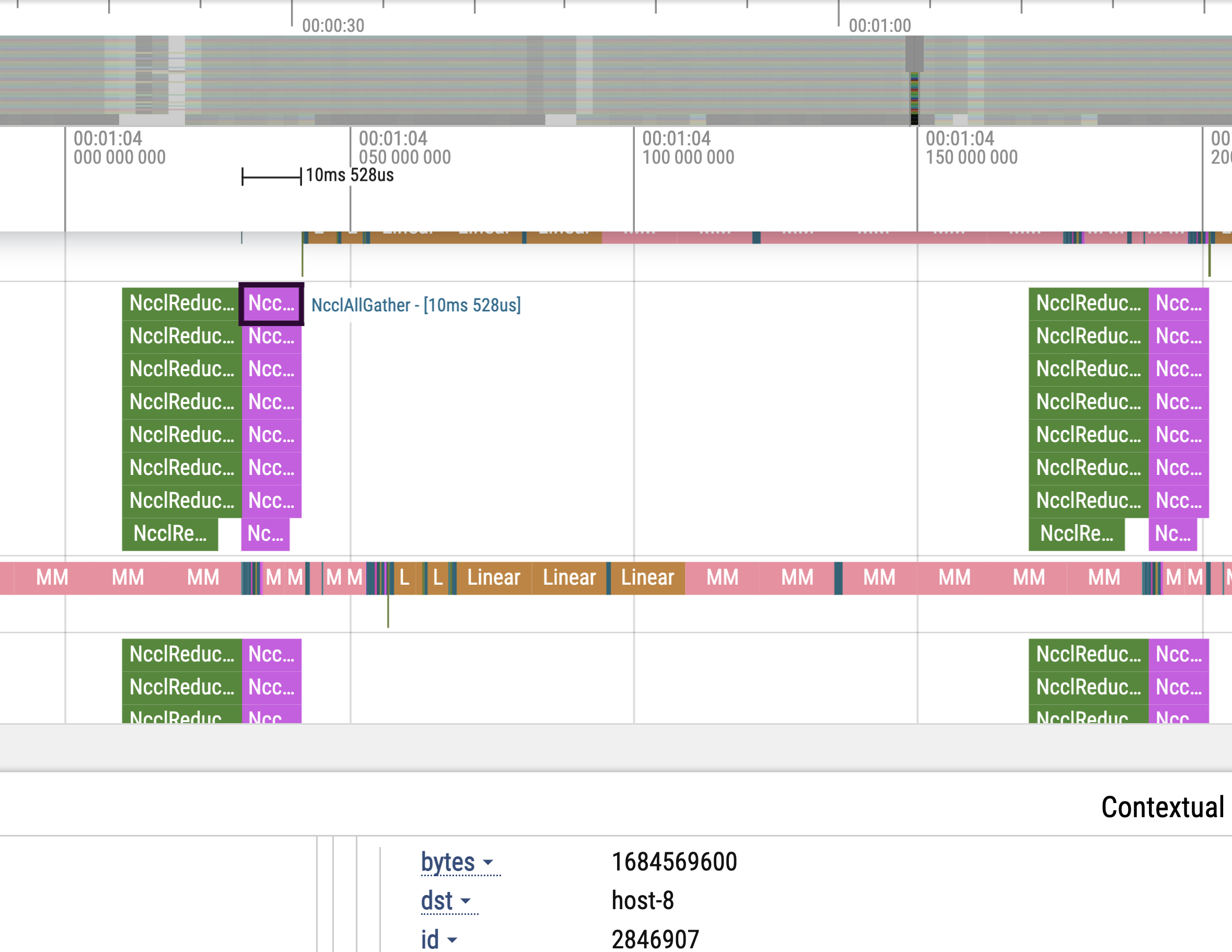}
\caption{Perfetto~\cite{perfetto} trace exported by \sys.}
\vspace{-3mm}
\label{fig:chrome-trace}
\end{figure}

\sys also supports feature-rich visualization via Perfetto UI~\cite{perfetto} to help developers tune their system. \autoref{fig:chrome-trace} shows a visualized simulation trace of TorchTitan exported by \sys.

\parab{Computation/communication overlap and dynamic memory behaviors in ML systems.} \sys naturally captures computation/communication overlap and other runtime behaviors in ML systems. \autoref{fig:chrome-trace} shows the timeline of \sys executing TorchTitan, where x-axis is simulated time. As shown in the figure, the NCCL operations (communication) overlap with matrix multiplication (computation).

\sys can also naturally capture dynamic behaviors of the PyTorch caching allocator as it tracks memory management on CUDA Runtime level.
Note that ML systems usually cannot utilize all of GPU memory due to memory fragmentation. \sys can precisely reflect the fragmentation and dynamic behaviors of the PyTorch caching allocator, leaving the only imprecision under CUDA Runtime, i.e., the memory management in NVIDIA GPU driver.

\subsection{Simulation Accuracy}

\begin{figure*}[t]
    \centering
    \vspace{1mm}
    \includegraphics[width=.96\textwidth]{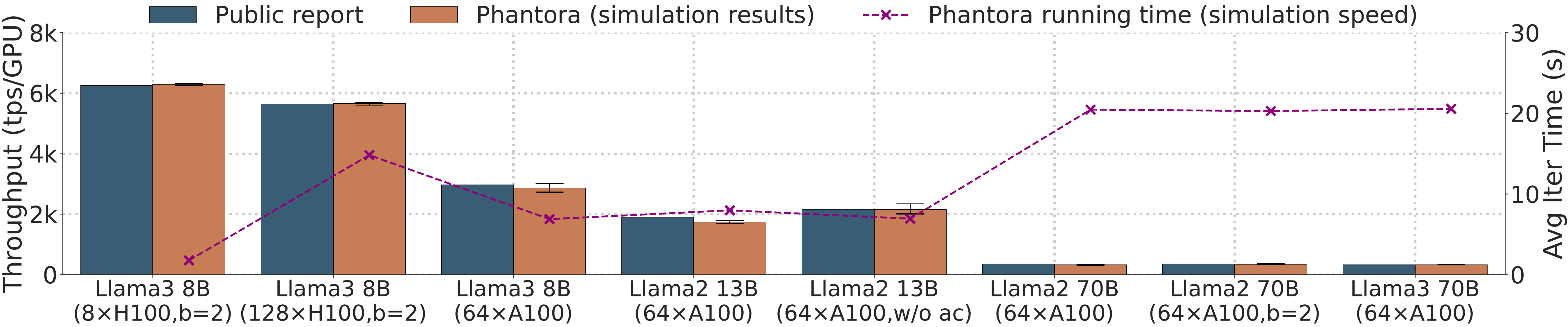}
    \caption{\textbf{Accuracy and speed of \sys (large scale):} Training throughput reported by TorchTitan~\cite{liang2025torchtitan} using FSDP2, simulation results and simulation speed of \sys on the testbeds. The error bars show 95\% confidence interval. ``ac'' means activation checkpointing in TorchTitan~\cite{liang2025torchtitan}. 
    }
    \label{fig:eval-accuracy}
\end{figure*}

\parab{Our hardware testbed.} We run \sys on two on-premise GPU servers, depending on which server has the closer hardware to the performance report we are comparing with.
One server is equipped with 2 AMD EPYC 9355 CPUs and 4 NVIDIA H200 NVL GPUs connected via NVLink.
We use all four GPUs to collect groud truth performance numbers. When we run \sys, we restrict the GPU usage to a single GPU. We use \textit{H200 testbed} to reference this testbed.
Another server is equipped with 2 Intel Xeon Gold 6348 CPUs and a single NVIDIA A100 40G GPU.
We use \textit{A100 testbed} to reference this testbed.

\begin{figure}[t]
    \centering
    \includegraphics[width=.47\textwidth]{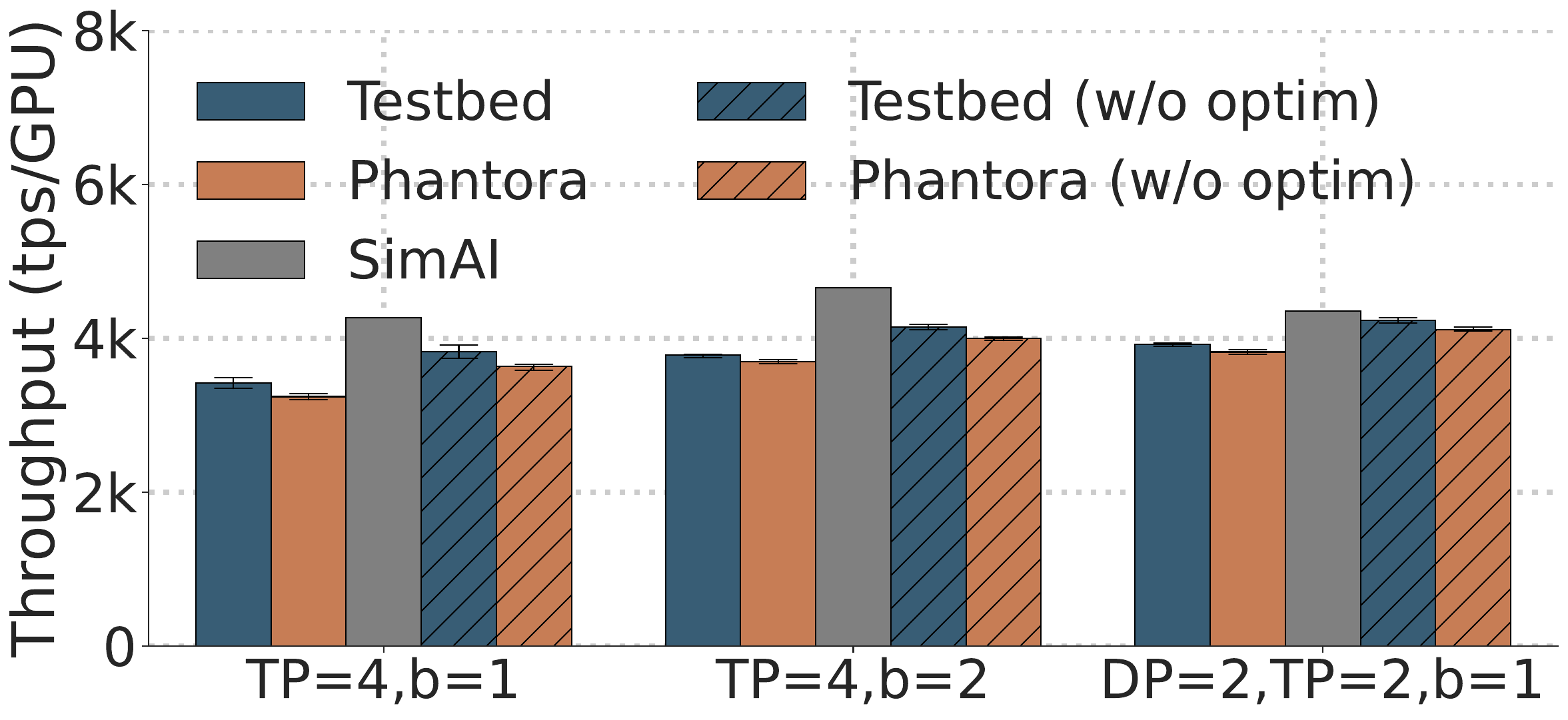}
    \caption{\textbf{Accuracy of \sys (small scale):} Megatron training throughput of Llama2 7B on H200 testbed with or without optimizer, \sys simulation results with or without optimizer, and SimAI simulation results. The error bars show 95\% confidence interval. Note that SimAI currently does not include optimizer in its simulation.}
    \label{fig:eval-accuracy-small}
\end{figure}

\parab{Comparing with public performance report as ground truth.} TorchTitan provides a comprehensive benchmark results~\cite{torchtitana100,liang2025torchtitan} utilizing up to 128 GPUs with a combination of FSDP2 and activation checkpointing. \autoref{fig:eval-accuracy} shows the accuracy of \sys to simulate TorchTitan's benchmark.
The average error is 2.9\% with the maximum error of 8.5\% on Llama2 13B.
Note that due to the limitation of accessible hardware, H100 reports~\cite{liang2025torchtitan} are evaluated on the H200 testbed, and A100-80G reports~\cite{torchtitana100} are evaluated on the A100 testbed with a single A100-40G GPU. The difference of computing power (FLOPS) between the reported GPUs and testbed GPUs is minor. The main difference is memory capacity, which is configurable in \sys and is set to the corresponding amount (80GB) in both experiments.

\parab{Comparing with testbed training performance as ground truth.} \autoref{fig:eval-accuracy-small} shows the accuracy of \sys using Llama2 7B training with different parallelization strategies and batch sizes. Compared with the ground truth, the average error of \sys is 3.7\% with the maximum error of 5.3\% when tensor parallel size is 4 and micro batch size is 1. We hypothesize that SimAI’s error is larger than expected because a core component, SimCCL, though open-sourced, had not yet been integrated into SimAI’s open-sourced mocked frameworks at the time of the experiment.

\subsection{Simulation Speed and Scalability}
\label{sec:eval-scalability}

\begin{figure}[t]
    \centering
    \includegraphics[width=.46\textwidth]{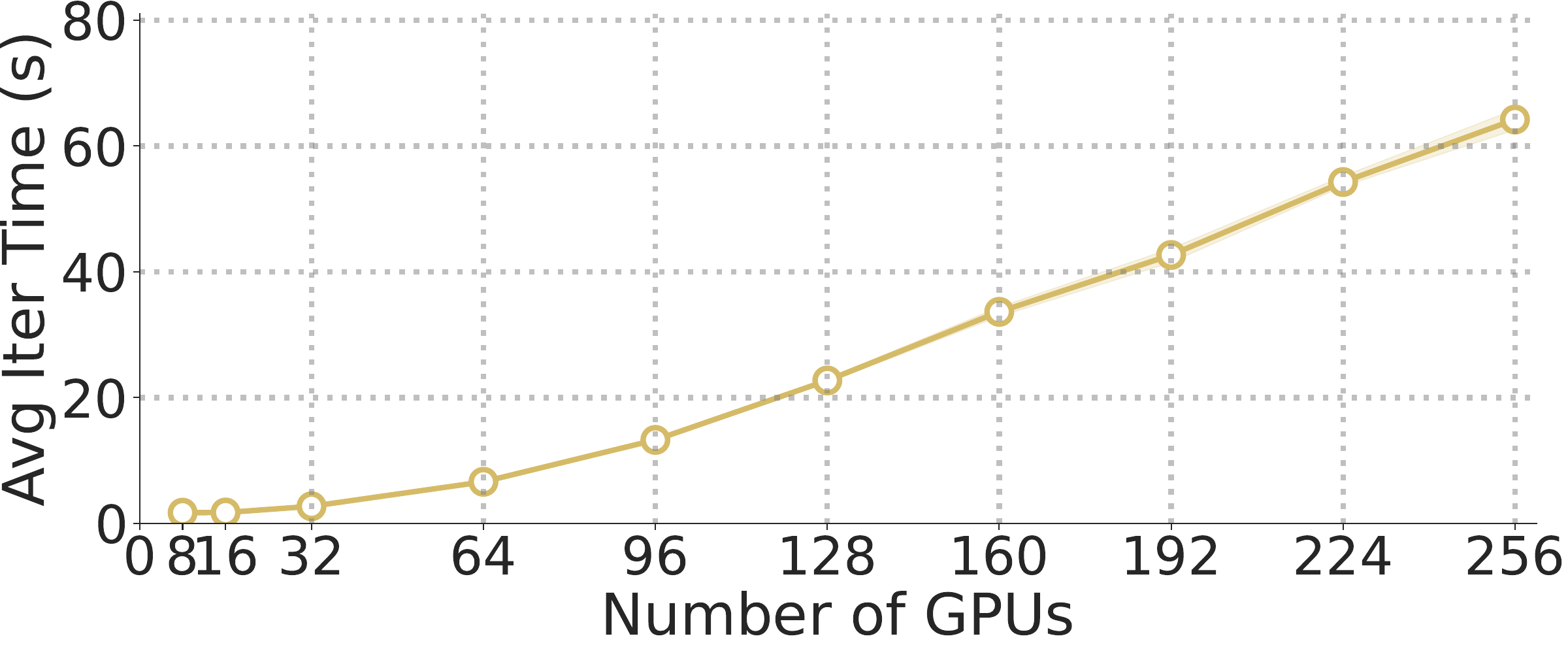}
    \caption{\textbf{Speed of \sys:} \sys simulation time of Llama2 7B training (Megatron, TP=8) using 32 CPU cores on H200 testbed.}
    \label{fig:eval-speed-large}
\end{figure}

\parab{Simulation speed.} \autoref{fig:eval-accuracy} also shows the simulation speed of \sys on TorchTitan's benchmark, which demonstrates \sys's ability to quickly evaluate a large scale workload. For example, Llama3 8B training with 128 GPUs takes around 15 seconds per iteration to simulate using \sys, which means the user can easily estimate the training throughput of this workload in minutes.

\autoref{tbl:eval-speed-small} shows the simulation speed of \sys in an actual training workload on the H200 testbed. The simulation time remains at the same level as the actual training time, and is significantly shorter than SimAI. This difference is mainly because \sys uses a flow-level network simulator, while SimAI uses a packet-level network simulator.

\autoref{fig:eval-speed-large} shows the \sys's simulation speed on Llama2 7B training using Megatron with different numbers of GPUs. The training uses data parallelism over tensor parallelism, where tensor parallel size is fixed to 8. The batch size per GPU is fixed to 1. \autoref{fig:eval-speed-large} shows that the simulation time increases linearly with respect to the workload scale when the number of GPUs is greater than 100, which is expected given fixed 32 CPU cores. If we set a limit of 1 minute per iteration, \sys can simulate approximately 240 GPUs under this setting. Note that \sys simulator is still single-threaded with one dedicated CPU core, and Megatron containers are placed on the other 31 cores.

\begin{table}[t]
\centering
\begin{tabular}{cccccc}
\toprule
DP & TP & batch & Testbed & \sys & SimAI \\
\midrule
1 & 4 & 1 & 0.30s & 0.91s & 56.9s \\
1 & 4 & 2 & 0.54s & 0.93s & 63.4s \\
2 & 2 & 1 & 0.52s & 0.96s & 117.7s \\
\bottomrule
\end{tabular}
\caption{\textbf{Speed of \sys (small scale):} Average time per iteration of Llama2 7B training using Megatron on H200 testbed; Average \sys running time per iteration and SimAI running time for one iteration. Both \sys and SimAI can use at most 16 cores. Note that SimAI uses packet-level network simulation while \sys uses flow-level network simulation.}
\vspace{-3mm}
\label{tbl:eval-speed-small}
\end{table}

\begin{figure}[t]
    \centering
    \includegraphics[width=.46\textwidth]{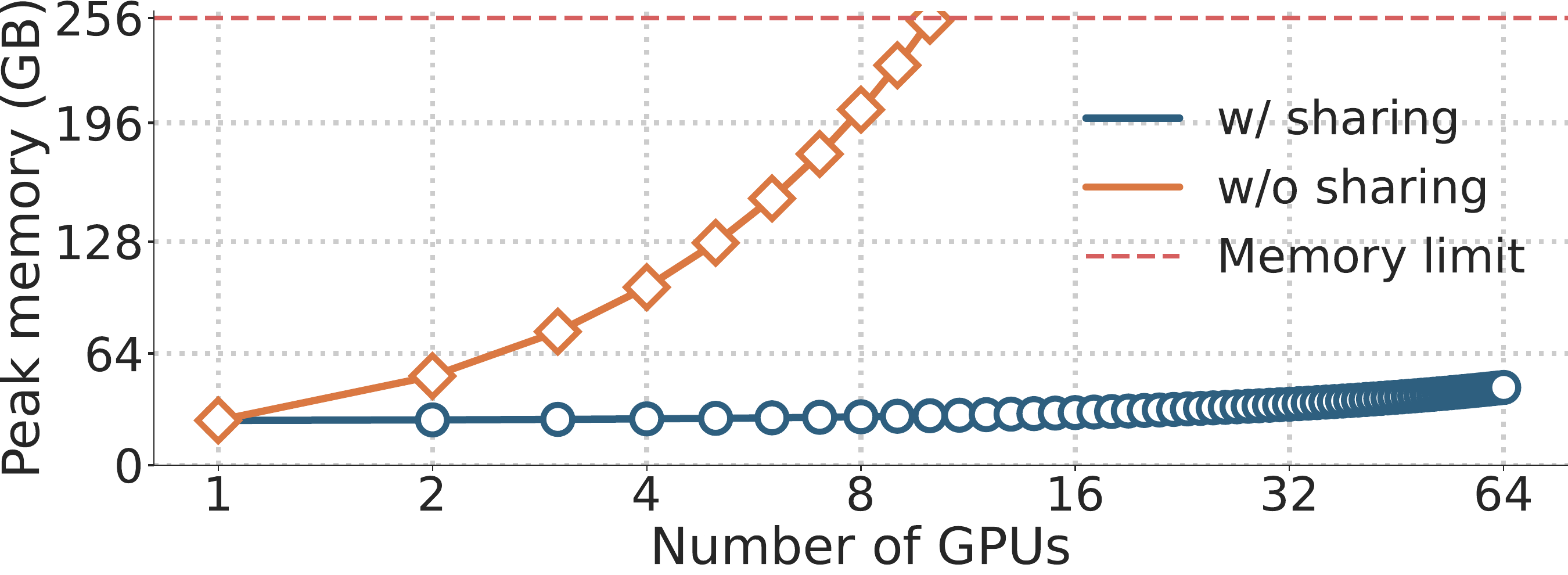}
    \caption{\textbf{CPU memory usage:} Peak CPU memory usage of Llama2 7B training simulation in DeepSpeed with or without model parameter sharing.}
    \vspace{-2mm}
    \label{fig:eval-memory-usage}
\end{figure}

\parab{CPU memory usage.} Sometimes users may choose to load or initialize a full model instead of a corresponding shard on each rank, especially when using a framework like DeepSpeed~\cite{rasley2020deepspeed}, which transparently and automatically shard all models. In this case, as discussed in \autoref{sec:scalability}, \sys implemented model parameter sharing on CPU memory to overcome scalability limitation. \autoref{fig:eval-memory-usage} shows the peak memory usage over different simulation scales: Without parameter sharing, 256GB of CPU memory can only support 9 GPUs, while 64 GPUs only need less than 64GB of memory with parameter sharing. This greatly improves the scalability of \sys without forcing users to change the training script.

\begin{figure*}[t]
    \centering
    \includegraphics[width=\textwidth]{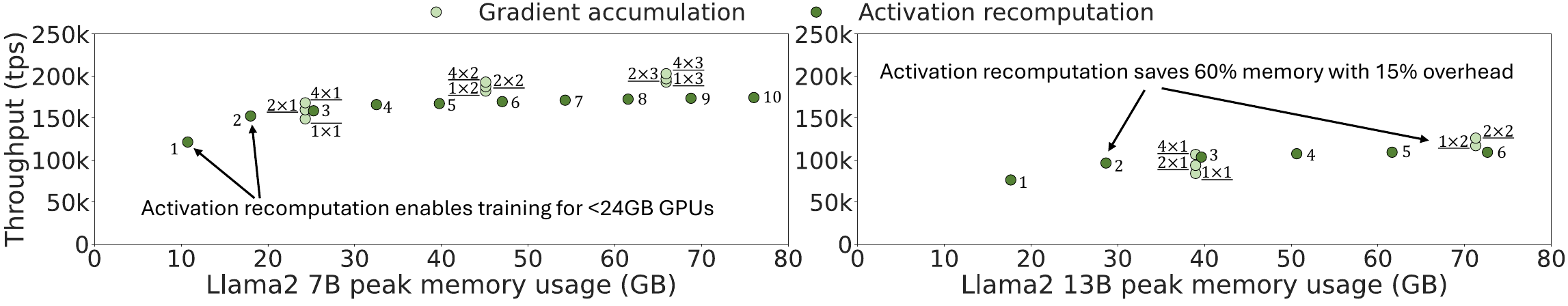}
    \caption{\textbf{Evaluating memory saving techniques:} \sys estimated peak memory usage per GPU and throughput of Llama2 training with different memory saving techniques (64 H100 GPUs, Megatron, DP=8, TP=8). A single number $n$ denotes the number of batches per GPU with activation recomputation. A pair in the form $m \times n$ denotes $n$ batches per GPU with $m$ gradient accumulation steps.}
    \vspace{-4mm}
    \label{fig:case_study}
\end{figure*}

\subsection{Case Study: Activation Recomputation}

Selective activation recomputation~\cite{korthikanti2023ac} is a technique to save GPU memory in large scale training by discarding certain intermediate activations in the forward pass and recomputing them in the backward pass. It is widely supported in many ML frameworks like Megatron~\cite{shoeybi2019megatron}, DeepSpeed~\cite{rasley2020deepspeed} and TorchTitan~\cite{liang2025torchtitan}. However, to the best of our knowledge, no other existing simulator can simultaneously analyze its effect on throughput and memory usage, because the existing simulators have not yet fully reimplemented this feature. Meanwhile, \sys can natively support this feature without implementing any specific logic related to selective activation recomputation.

\autoref{fig:case_study} shows the simulated peak memory usage and throughput of Llama2 training on 64 H100 GPUs, comparing activation recomputation with normal training with gradient accumulation---a more commonly used technique to achieve larger global batch size without increasing micro batch size.
Our simulation results largely align with the results in the original selective activation recomputation paper~\cite{korthikanti2023ac}: \textit{selective activation recomputation greatly reduces memory usage without introducing significant throughput overheads.} This experiment has shown that \sys can be used to assess the performance of new training optimizations without any redundant reimplementation effort for simulator developers.

\section{Discussion}
\label{sec:discussion}

\parab{General support of custom kernels.} As detailed in \autoref{sec:workflow}, \sys can support custom kernel extensions in an ad-hoc manner but still requires extra engineering effort. And like all existing workload-based performance estimation methods, \sys does not currently support JIT-compiled kernels. In principle, a more general and transparent support could be implemented by intercepting or cooperating with compilers, but we leave this direction for future work.

\parab{Value-dependent performance in ML frameworks.}
Like other ML system simulators, \sys currently cannot precisely reflect some value-dependent performance characteristics. One key example is expert parallelism~\cite{lepikhin2021gshard}, where performance depends on the distributions of activated experts. \sys can simulate expert parallelism under the assumption of perfect load balance, but it does not model the performance overheads caused by expert imbalance. Another example is reinforcement learning (RL) for LLMs, where the its performance depends on the generation length.
We believe this limitation can be addressed through an annotation interface that allows users to specify distributions of certain values (\eg, activated expert indices, LLM generation lengths). With these distributions, \sys can allow users to further estimate the performance of their system under different scenarios. We leave this direction to future work.

GPU itself can also has value-dependent performance of a single operation—for instance, sorting implementations on GPU may take different branches based on comparison of tensor values. We ignore such GPU-level effects as their performance impact is generally minor in end-to-end ML systems.

\parab{Improving \sys's accuracy.}
Our main goal is maximizing ML framework code reuse, not to improve simulation accuracy beyond existing static workload simulations.
Similar to static workload simulations, \sys uses standard discrete event simulation, so other techniques improving simulation accuracy should also apply to \sys. For example, the SimCCL library in SimAI~\cite{wang2025simai} could replace the current simple ring algorithm in \sys for more accurate network simulation of NCCL.

\parab{Improving \sys's speed.}
Similarly, \sys can adopt simulation performance enhancement techniques in existing static workload simulations. For example, while \sys simulator is currently single-threaded, parallel discrete event simulation like UNISON~\cite{bai2024unison} could potentially be applied to \sys for improved simulation speed.

\parab{Non-independent computation/communication overlap performance.}
Overlapping communication with computation is a standard way to hide communication latency in distributed ML systems. However, this overlap could also slow down both operations as they share critical internal hardware resources~\cite{liu2024deepseek}. Currently \sys and other simulators do not consider this effect as it's hard to simulate and impractical to profile for every possible overlap. An analytical model could help in the estimation, but we leave it for future work.

\section{Related Works}

\parab{Hybrid simulation.} We are not the first to consider integrating an event-driven simulator with a real system. In the 90s, the Wisconsin Wind Tunnel (WWT)~\cite{reinhardt1993wwt, mukherjee2000wwt} explored an execution-driven simulation approach to estimate performance of cache-coherent shared memory systems. WWT also encountered the problem of synchronizing simulated time with direct-execution. The approach adopted in WWT is to simulate every quantum, which is calculated from the minimum inter-core communication latency. In theory, we could also build \sys using this idea by carefully controlling the execution of containers (using interrupts to stop/resume container execution), so that containers' times are synchronized with the simulator. However, this approach would make \sys significantly slower. Another key difference between \sys and execution-driven simulation from the computer architecture community~\cite{martin2005gems} is that \sys does not faithfuly execute data operations. \sys does not track the GPU buffer content and does not execute GPU computation (except profiling runtime) or communication.

Another effort is to enable today's event-driven network simulator with real systems. For instance, ns-3~\cite{ns3} supports a feature called TapBridge, which implements a special Linux network device. This allows Linux applications to run over an ns-3 simulated network. However, this approach introduces inaccuracies in performance estimation. ns-3 does not control the system time of the Linux environment in the same way \sys controls the time of the ML system. As a result, if ns-3 takes 1 second to simulate 10 ms of network activity, the Linux application perceives the network as 100 times slower than it actually is. 

\parab{Predicting CUDA kernel execution time.} Predicting a CUDA kernel's execution time is a standard problem in ML compilers~\cite{chen2018tvm, zheng2020ansor}. The reason is that an ML compiler's goal is to generate the most efficient CUDA kernel for ML operations.
The approaches compilers take are usually ML-based performance modeling, which eliminates the need to test every CUDA kernel's performance on real hardware. For example, TVM \cite{chen2018tvm} trains a gradient boosting decision tree by applying a set of manually-designed features on generated code. However, our problem of predicting CUDA kernel performance differs significantly from that of ML compilers. We focus on a limited set of kernels---those already selected by the ML systems (\eg, from PyTorch, Megatron \cite{shoeybi2019megatron}, DeepSpeed \cite{rasley2020deepspeed} and TorchTitan \cite{liang2025torchtitan}).
So, \sys can use comprehensive runtime profiling of these CUDA kernels instead of other methods to estimate their performance.

\parab{Emulating the control plane and simulating the data plane.} We draw inspirations from CrystalNet \cite{liu2017crystalnet} for network testing. CrystalNet's idea is to emulate a network environment for switch control programs without actually forwarding data. This allows CrystalNet to fully emulate large production networks on only tens of VMs/containers to find network bugs.
Similarly, in \sys, all the ML system code are running as is except the GPU operations and communication are simulated. Although in different contexts, both CrystalNet and \sys rely on the assumption that the control flow does not have data dependency.

\section{Conclusion}
This paper introduces \sys, a hybrid GPU cluster simulator for ML system performance estimation. \sys runs unmodified ML models and frameworks, intercepting and simulating GPU- and network-related operations for high-fidelity performance estimation. It addresses key research challenges, including supporting unmodified ML frameworks, the integration of event-driven network simulators with real-time code execution, along with techniques to improve simulation scalability. Our evaluation shows that, \sys achieves similar estimation accuracy to state-of-the-art static workload simulation methods. At the same time, \sys's design is general, supporting three state-of-art LLM training frameworks out-of-box. Further, \sys only requires a single GPU, eliminating the need for the resource-intensive trace collection and workload extraction steps required by traditional trace-based simulators.

\section*{Acknowledgements}
This work was supported in part by NSF grants CNS-2112562, CNS-2238665, CNS-2330333, CNS-2402696, and OAC-2503010, as well as by gifts from Amazon and Meta.
This work was also supported in part by the Center for Ubiquitous Connectivity (CUbiC), sponsored by Semiconductor Research Corporation (SRC) and Defense Advanced Research Projects Agency (DARPA) under the JUMP 2.0 program.

\bibliographystyle{plain}
\bibliography{confs_long, ref}

\appendix

\section{Evaluation for Non-LLM Workloads}
\label{sec:nonllm}

\sys's design does not depend on any particular model architectures. Here we evaluate \sys for non-LLM workloads.

We use a single NVIDIA RTX 3090 GPU for computation kernel profiling. The goal is to match the actual performance of our 4-server testbed where each server has 2 Intel Xeon Gold 5215 CPUs and 2 NVIDIA RTX 3090 GPUs. There are 8 NVIDIA RTX 3090 GPUs in total. The network topology and bandwidth of this testbed is the input to our network simulator. We evaluate a broad set of models on DeepSpeed~\cite{rasley2020deepspeed} that are known to have different performance characteristics, including ResNet-50 \cite{he2016deep}, Stable Diffusion~\cite{stable_diffusion}, 
and Graph Attention Network (GAT)~\cite{gat}.

\begin{figure}[t]
    \centering
    \includegraphics[width=0.45\textwidth]{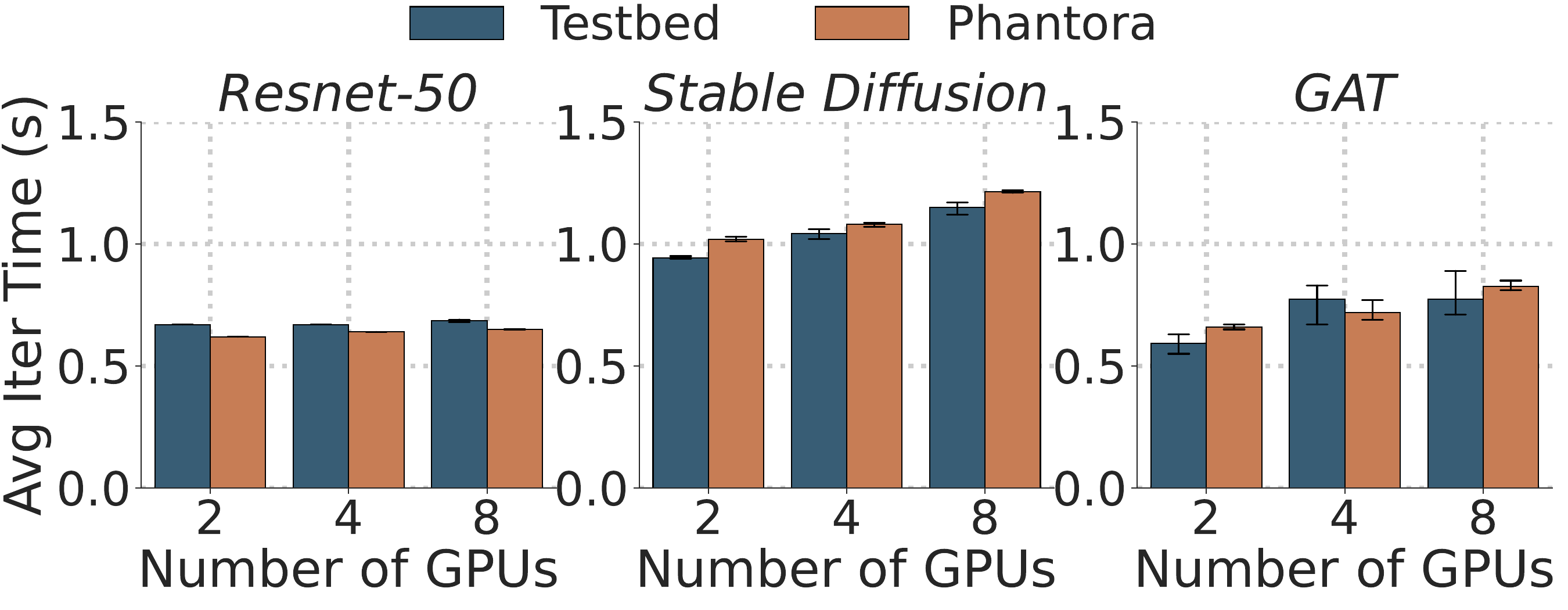}
    \caption{\textbf{Non-LLM models:} Testbed training time and \sys's simulation results with DeepSpeed. The error bars show 95\% confidence interval.
    }
    \label{fig:eval-smallmodels}
\end{figure}

\sys achieves accurate simulation accuracy. \autoref{fig:eval-smallmodels} compares \sys's predicted training time per iteration and grouth truth performance numbers collected on the testbed. the average simulation error is 6.6\% under these settings with the maximum error of 8.1\% on diffusion model with 2 GPUs.

\end{document}